\documentclass[12pt,thmsa]{article}
\usepackage{amssymb}
\usepackage{graphicx}

\input{tcilatex}
\begin{document}

\begin{center}
{\large Synthesis of maximally entangled mixed states and
disentanglement in coupled Josephson charge qubits}

\textbf{M. Abdel-Aty}\footnote{%
E-mail: abdelatyquant@yahoo.co.uk}

{\footnotesize
Mathematics Department, Faculty of Science, South
Valley University, 82524 Sohag, Egypt \\ Mathematics Department,
College of Science, Bahrain University, 32038 Kingdom of Bahrain
}

~

{\bf Eur. Phys. J. D Vol. 46, pp. 537-543 (2008)}

\end{center}

We analyze a controllable generation of maximally entangled mixed
states of a circuit containing two-coupled superconducting charge
qubits. Each qubit is based on a Cooper pair box connected to a
reservoir electrode through a Josephson junction. Illustrative
variational calculations were performed to demonstrate the effect on
the two-qubits entanglement. At sufficiently deviation between the
Josephson energies of the qubits and/or strong coupling regime,
maximally entangled mixed states at certain instances of time is
synthesized. We show that entanglement has an interesting subsequent
time evolution, including the sudden death effect. This enables us
to completely characterize the phenomenon of entanglement sharing in
the coupling of two superconducting charge qubits, a system of both
theoretical and experimental interest.

\[
\]

\section{Introduction}

There have been remarkable advances in the quest to build a
superconductor-based quantum information processor in recent years
\cite{nis07,you05} and one of the greatest scientific and
engineering challenges of this decade is the realization of a
quantum computer. In this context, a solid-state system is highly
desirable because of its compactness, scalability and
compatibility with existing semiconductor technology. One of the
physical realizations of a solid-state qubit is provided by a
Cooper pair box which is a small superconducting island connected
to a large superconducting electrode, a reservoir, through a
Josephson junction \cite{gri07}. Superconducting charge qubits
(Cooper pair boxes) are a promising technology for the realization
of quantum computation on a large scale
\cite{coo04,nak99,pas05,pas03,ast06}.

Using simultaneous measurement and state tomography, entanglement
between two solid-state qubits has been demonstrated \cite{ste06}.
The results demonstrate a high degree of unitary control of the
system, indicating that larger implementations are within reach.
These results are promising for future solid-state quantum
computing. For conventional fault-tolerant quantum computing, the
quantum states should have a high level of purity, preferably
being as close to a pure state as possible. When the qubit is
coupled to an environment it is subject to decoherence, which will
typically result in a completely mixed state \cite{you02}.
However, a qubit initially in a completely mixed state can be
purified by measurement. Therefore, it is desirable on both
fundamental and practical grounds to study maximin entangled state
generation and entanglement dynamics in a time-dependent sense.
One of the next major steps towards building a Josephson junction
quantum computer prototype will be the demonstration of
controllable coupling between the qubits. At this end, it seems
that a quantitative link between the degree of disentanglement and
the amount of the energy transferred between the system of
interest and its environment is still missing. Investigation of
entanglement control in such systems would therefore be an
important contribution to the present suite of experimental
controls.

In recent years quantum entanglement has found many exciting
applications that have considerable bearing on the emerging fields
of quantum information and quantum computing
\cite{nie00,yu07,ebe07,alm07,lou05}. Moreover, besides this
fundamental aspect, the interest in entangled states has been
recently renewed because their properties lie at the heart of many
potential applications. The generation and reconstruction of
quantum states were extensively studied in the past theoretically
and experimentally \cite{zha07,zou05,sch97,mun01,di04}. More
fundamentally, decoherence processes due to the interaction with
internal or external noises and entanglement decay in a
time-dependent sense have been studied in many distinct cases
\cite{yu07,dio03,yu02,yu03,yu04,tol05,aty07}. More recently,
Almeida et al. \cite{alm07} have devised an elegantly clean way to
confirm the existence of so-called entanglement sudden death, that
is, entanglement terminates completely after a finite interval,
without a smoothly diminishing long-time tail.

This paper examines the generation of maximally entangled mixed
state of two-coupled Josephson charge qubits using a common pulse
gate. We present various examples in order to monitor different
regimes of synthizing the maximally entangled mixed states and
entanglement dynamics. In principle, by proper adjustment of the
initial state parameters, we can always find suitable values of
characteristic energies of the Cooper pairs and coupling energy
which can be used to suppress the decay of entanglement. This
analysis is carried out with generalized time-dependent density
matrix, yielding a generalized dynamical two-qubit model. As we
are going to show, we may take this advantage to increase time
intervals for maximally entangled states caused by the strong
coupling regime.

This paper is organized as follows: In Sec. 2, we will describe
the Hamiltonian of the system of interest, and obtain the explicit
analytical solution of the master equation describing the dynamics
of two qubits in the presence of phase decoherence. In Sec. 3, by
calculating the occupation probabilities of the two qubits, we
show that it is possible to generate the maximally entangled mixed
states of the system in different situations. In Sec. 4, we
discuss the entanglement of the system by virtue of the
concurrence in the absence or presence of the decoherence.
Finally, Sec. 5 presents the conclusions and an outlook.

\begin{figure}[h]
\begin{center}
\includegraphics[width=11cm]{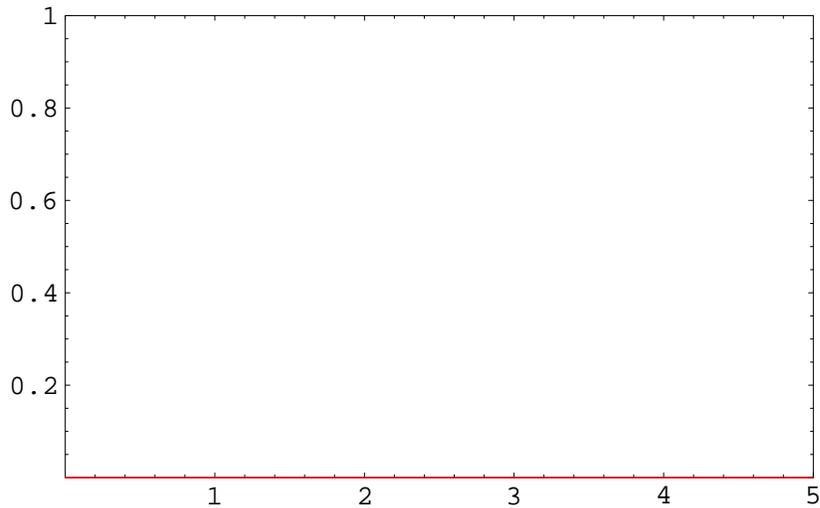}
\end{center}
\caption{Illustration of two capacitively coupled Josephson charge
qubits. The circuite consists of two charge qubits that are
coupled by an on-chip capacitor $C_{m}$ \protect\cite{pas05,hu07}.
} \label{sch}
\end{figure}

\section{Two coupled charge qubits}

Here, we briefly discuss the general formalism to characterize the
dynamics of two-coupled superconducting charge qubits (Cooper pair
boxes connected to a reservoir electrode through a Josephson
junction). For a more detailed discussion we refer the reader to
Ref. \cite{pas05,ave85}. We consider two charge qubits and couple
them by means of a miniature on-chip capacitor. The read-out of
each qubit, in this case, is done similar to the single qubit
read-out and connect a probe electrode to each qubit. External
controls that
we have in the circuit are the dc probe voltages $V_{b_{1}}$ and $V_{b_{2}}$%
, $d_{c}$ gate voltages $V_{g_{1}}$ and $V_{g_{2}}$, and pulse gate voltage $%
V_{p}$ (see figure 1). The information on the final states of the
qubits after manipulation comes from the pulse-induced currents
measured in the probes. By doing routine current-- voltage--gate
voltage measurements, we can estimate the capacitances. We then
perform state manipulation and demonstrate qubit--qubit
interaction. The Hamiltonian of the system in the
charge representation can be written as%
\begin{equation}
\hat{H}=\hbar \sum_{n_{1}=0}^{\infty }\sum_{n_{2}=0}^{\infty }\eta
_{1}(n_{1},n_{2})\widehat{S}_{11}-\frac{E_{J_{1}}}{2}\left( \widehat{S}_{12}+%
\widehat{S}_{34}\right) -\frac{E_{J_{2}}}{2}\left( \widehat{S}_{13}+\widehat{%
S}_{24}\right) ,  \label{ham1}
\end{equation}%
where $\widehat{S}_{11}=|n_{1},n_{2}\rangle \langle n_{1},n_{2}|,$ $\widehat{%
S}_{12}=|n_{1},n_{2}\rangle \langle n_{1}+1,n_{2}|,$ $\widehat{S}%
_{13}=|n_{1},n_{2}\rangle \langle n_{1},n_{2}+1|,$ $\widehat{S}%
_{34}=|n_{1},n_{2}+1\rangle \langle n_{1}+1,n_{2}+1|,$ and $\widehat{S}%
_{24}=|n_{1}+1,n_{2}\rangle \langle n_{1}+1,n_{2}+1|.$ The
parameter $\eta _{1}(n_{1},n_{2})=E_{c_{1}}\left(
n_{g_{1}}-n_{1}\right) ^{2}+E_{c_{2}}\left( n_{g_{2}}-n_{2}\right)
^{2}+E_{m}\left(
n_{g_{1}}-n_{1}\right) \left( n_{g_{2}}-n_{2}\right) .$ Here, $n_{1}$ and $%
n_{2}$ ($n_{1},n_{2}=0,\pm 1,\pm 2,...$) are the numbers of excess
Cooper
pairs in the first and the second Cooper pair boxes, and $%
n_{g_{1,2}}=(C_{g_{1,2}}V_{g_{1,2}}+C_{p}V_{p})/2e$ are the
normalized charges induced on the corresponding qubit by the
$d_{c}$ and pulse gate
electrodes. The eigenenergies, $E_{k}$ $(k=0,1,2,...$), of the Hamiltonian (%
\ref{ham1}) form $2e$-periodic energy bands corresponding to the
ground ($k=0 $), first excited ($k=1$), etc. states of the system.
$E_{c_{1}}$, $E_{c_{2}} $ and $E_{m}$ give the characteristic
energies of Cooper pair of the first qubit, Cooper pair charging
energy of the second qubit and the coupling energy, respectively.
\begin{eqnarray}
E_{c1,2} &=&\frac{4e^{2}C_{\varepsilon 2,1}}{2(C_{\varepsilon
1}C_{\varepsilon 2}-C_{m}^{2})},  \nonumber \\
E_{m} &=&\frac{4e^{2}C_{m}}{C_{\varepsilon 1}C_{\varepsilon
2}-C_{m}^{2}},
\end{eqnarray}%
where $C_{\varepsilon _{1,2}}$ are the sum of all capacitances
connected to the corresponding Cooper pair box including the
coupling capacitance $C_{m}$ and $e$ is the electron charge.

If the circuit is fabricated to have the following relation
between the characteristic energies: $E_{J_{1,2}}\sim $
$E_{m}<E_{c_{1,2}}$, then one
can use a four-level approximation for the description of the system ($%
|00\rangle ,|01\rangle ,|01\rangle $ and $|11\rangle $) around $%
n_{g_{1}}=n_{g_{2}}=0.5$ while other charge states are separated
by large energy gaps. In this basis, the two charge qubits system
behaves as a single
four-level system which can be used as a new basis for the Hamiltonian (\ref%
{ham1}).

The time evolution of the system density operator $\hat{\rho}(t)$
can be written as \cite{bre02,lid03,gar00}
\begin{equation}
\frac{d}{dt}\hat{\rho}(t)=-\frac{i}{\hbar
}[\hat{H},\hat{\rho}]-\frac{\gamma }{2\hbar
^{2}}[\hat{H},[\hat{H},\hat{\rho}]],  \label{mas}
\end{equation}%
where $\gamma $ is the phase decoherence rate. Equation
(\ref{mas}) reduces
to the ordinary von Neumann equation for the density operator in the limit $%
\gamma \rightarrow 0.$ The equation with the similar form has been
proposed to describe the intrinsic decoherence \cite{mil91}. Under
Markov approximations the \ solution of the master equation can be
expressed in
terms of Kraus operators \cite{yu06} as follows%
\begin{eqnarray}
\hat{\rho}(t) &=&\sum_{m=0}^{\infty }\frac{\left( \gamma t\right) ^{m}}{m!}%
\hat{H}^{m}\exp \left( -i\hat{H}t\right) \exp \left( -\frac{\gamma t}{2}\hat{%
H}^{2}\right) \hat{\rho}(0)\exp \left( -\frac{\gamma
t}{2}\hat{H}^{2}\right)
\exp \left( i\hat{H}t\right) \hat{H}^{m}  \nonumber \\
&=&\sum_{m=0}^{\infty }\frac{\left( \gamma t\right) ^{m}}{m!}\hat{M}^{m}(t)%
\hat{\rho}(0)\hat{M}^{\dagger m}(t),\qquad \qquad \qquad
\label{dens}
\end{eqnarray}%
where $\hat{\rho}(0)$ is the density operator of the initial state
of the system and $\hat{M}^{m}$ are the Kraus operators which
completely describe the reduced dynamics of the qubits system,
\begin{equation}
\hat{M}^{m}=\hat{H}^{m}\exp (-i\hat{H}t)\exp \left( -\frac{\gamma t}{2}\hat{H%
}^{2}\right) .
\end{equation}%
Equation (\ref{dens}) can also be written as
\begin{eqnarray}
\rho (t) &=&\exp \left( -i\hat{H}t\right) \exp \left( -\frac{\gamma t}{2}%
\hat{H}^{2}\right) \{e^{\hat{S}_{M}t}\hat{\rho}(0)\}\exp \left( -\frac{%
\gamma t}{2}\hat{H}^{2}\right) \exp \left( i\hat{H}t\right)
\label{dens2}
\\
&=&\rho _{ij,lk}(t)|ij\rangle \langle lk|.  \nonumber
\end{eqnarray}%
We define the superoperator $\hat{S}_{M}\hat{\rho}(0)=\gamma \hat{H}\hat{\rho%
}(0)\hat{H}$ and \textrm{choose arbitrary initial state of the two
charge qubits}$.$ The notation $|ij\rangle =|i\rangle _{1}\otimes
|j\rangle _{2},$ is used, where $|0\rangle _{1(2)}$ and $|0\rangle
_{1(2)}$ are the basis states of the first (second) qubits and
$\rho _{ij,lk}(t)=\langle ij|\rho (t)|lk\rangle $ corresponds the
diagonal ($ij=lk$) and off-diagonal ($ij\neq lk)$ elements of the
final state density matrix $\rho (t).$ From here on,
for tractability of notation and without loss of generality, we denote by $%
\rho _{ij}(t)=\rho _{ij,ij}(t),$ the probability of finding the
two-coupled charge qubits in the state $|ij\rangle .$

\section{Creation of maximally entangled mixed states}

We now apply the above results to study the time evolution of the
occupation probabilities with different values of the system
parameters. In pure-state case, there has been considerable debate
over the entanglement properties of certain types of states
\cite{nie00}. In this paper, we are interested in the case in
which the final state of the coupled charge qubits $\rho (t)$ is a
maximally entangled mixed state \cite{mun01}.

\begin{figure}[tbph]
\begin{center}
\includegraphics[width=11cm,height=5cm]{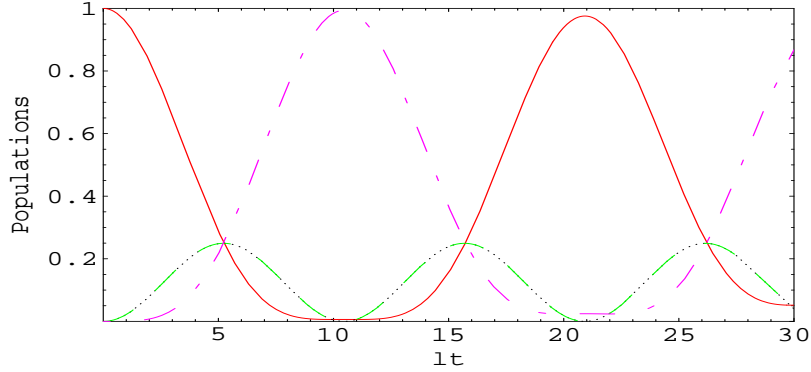}
\end{center}
\caption{The time evolution of the occupation probabilities $\protect\rho %
_{ij}(t),$ where $\protect\rho _{00}(t)$ (solid curve), $\protect\rho %
_{01}(t)$ (dashed curve), $\protect\rho _{10}(t)$ (dot-dashed curve) and $%
\protect\rho _{11}(t)$ (dotted curve). The initial state of the
two-charge qubits is assumed to be  $\rho(0)=|00\rangle\langle
00|$ and the parameters used in these figures are
$E_{J1}=E_{J2}=30$ and $E_{m}=6.$} \label{1}
\end{figure}
\begin{figure}[tbph]
\begin{center}
\includegraphics[width=11cm,height=5cm]{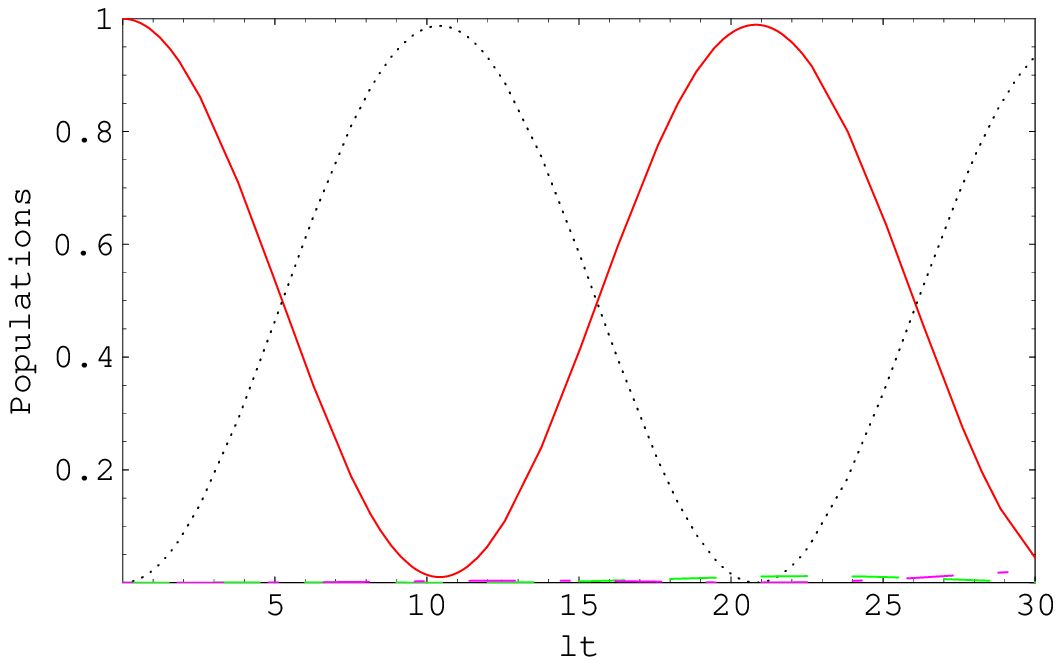}
\end{center}
\caption{{}The same as figure 1, but $E_{J1}=30\protect\mu eV,E_{J2}=5%
\protect\mu $eV and $E_{m}=6\protect\mu eV.$} \label{2}
\end{figure}
\begin{figure}[tbph]
\begin{center}
\includegraphics[width=11cm,height=5cm]{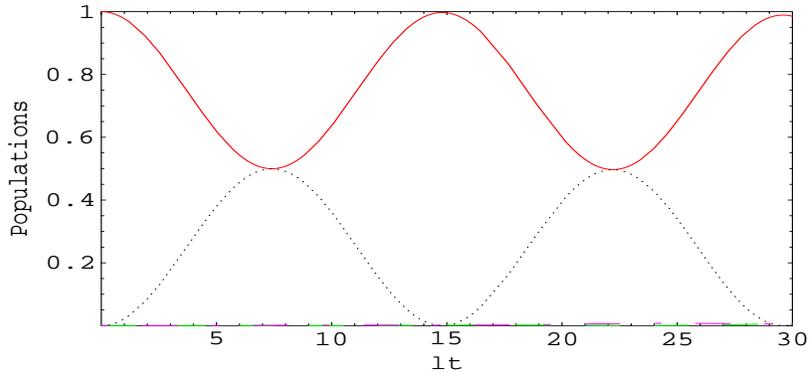}
\end{center}
\caption{{}The same as figure 1, but $E_{J1}=30\protect\mu eV,E_{J2}=1%
\protect\mu $eV and $E_{m}=60\protect\mu eV.$} \label{3}
\end{figure}

Nakamura et al. \cite{nak01} investigated the temporal behavior of
a Cooper-pair box driven by a strong microwave field and observed
the Rabi oscillations with multi-photon exchanges between the
two-level system and the microwave field. Here, the occupation
probabilities as functions of \ the scaled time $\lambda t$ are
schematically shown in Fig. \ref{1}. Note that the populations of
the four states exist but $\rho _{00}(t)$ as well as
$\rho _{10}(t)$ oscillate between $0$ and $1,$ while $\rho _{01}(t)$ and $%
\rho _{11}(t)$ oscillate with smaller amplitudes. It should be
noted that the occupation probabilities results are drastically
different when we consider different initial state settings. To
analyze the effect of the system parameters on the occupation
probabilities for the present system we consider two different
cases. One when the the Josephson energies of Cooper pair are
different while the second case is the strong interaction regime.
This will be seen in figure \ref{2} and figure \ref{3}. As an
example of the
creation of the two-particle maximally entangled state is shown in Fig. \ref%
{2}. The results of this figure are obtained for parameters $E_{J1}=30\mu $eV%
$,E_{J2}=5\mu $eV and $E_{m}=6\mu $eV. The way to determine
experimentally the qubits' Josephson energies $E_{J1}$ and
$E_{J2}$ has been described in Ref. \cite{ave85}

If the system starts from excited state, $\rho (0)=\left\vert
0,0\right\rangle \left\langle 0,0\right\vert ,$ we see that the
occupation probabilities of the intermediate states tend to zero
at any instant of time and the Cooper pair system oscillate only
between excited and ground states. If the environment is switched
off, i.e. $\gamma $ tends to zero, and using equation
(\ref{dens2}), at some instant times $\tau \simeq
\frac{5n}{\lambda },(n=1,2,3,...)$, we can obtain analytically the
values of the diagonal and off-diagonal elements of the density
matrix as $\rho _{00,00}(t)=\rho _{11,11}(t)=0.5$ and $\rho
_{00,11}(t)=\rho _{11,00}(t)=\zeta \neq 0,$ otherwise \ $\rho
_{ij,lk}(t)=0.$ Which means that, the final state takes the form

\begin{equation}
\hat{\rho}(t)=\frac{1}{2}\left\vert 0,0\right\rangle \left\langle
0,0\right\vert +\zeta (\left\vert 0,0\right\rangle \left\langle
1,1\right\vert +\left\vert 1,1\right\rangle \left\langle 0,0\right\vert )+%
\frac{1}{2}\left\vert 1,1\right\rangle \left\langle 1,1\right\vert
, \label{ana2}
\end{equation}%
i.e. the final state (\ref{ana2}) becomes a maximally entangled
mixed state \cite{mun01}. This entangled state corresponds to the
half-probability peaks in figure \ref{2} ($\zeta =0.13)$.
Depending on whether Josephson energy of the first qubit is
smaller or larger than the Josephson energy of the second qubit,
the maximally entangled states are created. Similarly, depending
on the coupling energy the maximally entangled state is
characterized as having short or long correlation time. Given
enough time, the system will therefore reaches a state where both
excited and ground states have equal occupation probabilities i.e.
the coupled-qubit system evolves to the maximally entangled mixed
state at the times given by

\begin{equation}
\lambda t=\frac{n\pi }{\sqrt{(E_{J1}+E_{J2})^{2}+E_{m}^{2}/4}-\sqrt{%
(E_{J1}-E_{J2})^{2}+E_{m}^{2}/4}},n=\pm 1,\pm 2,\pm 3,....
\end{equation}

We turn our attention to consider the case in which both
characteristic energies have different values taking into account
the effect of the coupling energy. For this reason we have plotted
the function $\rho _{ij}(t)$ against the scaled time $\lambda t$
in figures (\ref{3}). To make a comparison between this case and
the previous one we have to take the values of the other
parameters similar to that of the previous case. In this case
and providing the characteristic energies $E_{J1}=30\mu $eV$,$ and $%
E_{J2}=6\mu eV,$ we find that the function reduces its value to be around $%
\approx 1$ and $0.5$ for the excited state while the ground state
probability oscillates between $0$ and $0.5$. Furthermore, we
realize there is a long period of the interaction time in which
the probability of the excited state equals the ground state
probability, with perfect symmetric fluctuation pattern around
$0.5$, see figure (\ref{3}). Which means that, with these setting,
we obtained long lived maximally entangled mixed state (in this
case $\zeta =0.19)$. In the meantime if we exchange the values of
the characteristic energies$,$ we observe there is no big change
in the figure shape, except some decreases in the fluctuations
number. It is to be noted that the maximally entangled state in
this case is lived longer compared with significantly short time
in the previous case (see figures \ref{2} and \ref{3}).

One might now raise the following notes: taking a strong coupling
regime where the coupling energy is strong enough, one can obtains
maximally entangled mixed states at some instant times in a
periodical manner. But when the coupling energy between the two
qubits is weak, the period becomes shorter. Also, if the deviation
between the Josephson energies is substantially large, the
maximally entangled states can be generated. It is worth noting
that when the coupling energy tends to zero, our model becomes
similar to that of a beam splitter model. In such a case, we
cannot obtain maximally entangled states.

The above discussion clearly shows that the maximally entangled
state generation of the two charge qubits depends on both the time
evolution, Josephson energies of both charge qubits and coupling
energy. The considerations of experimental observability of the
entangling power discussed in \cite{gri07} are valid in the
context of the present work.

\section{Entanglement}

Having established the existence of maximally entangled states
$\rho (t),$ in section 3, now we try to answer the following
question: how does the entanglement of the two charge qubits
system evolve? To answer this question, one first needs a formal
definition of entanglement. Currently a variety of measures are
known for quantifying the degree of entanglement in a bipartite
system \cite{zho02,vid02,per96,lee00,zhe00,woo98,mes03}. A
convenient measure of entanglement for a two-qubit state $\rho
(t)$ is the concurrence $C_{\rho }\left( t\right) ,$ given by
\begin{equation}
C_{\rho }\left( t\right) =\max \left\{ 0,\lambda _{1}-\lambda
_{2}-\lambda _{3}-\lambda _{4}\right\} ,
\end{equation}%
where $\lambda _{1}\geq \lambda _{2}\geq \lambda _{3}\geq \lambda
_{4}.$ We denote by $\lambda _{i}$ the square roots of the
eigenvalues of $\rho \left( \sigma _{y}\otimes \sigma _{y}\right)
\left( \rho \right) ^{\ast }\left( \sigma _{y}\otimes \sigma
_{y}\right) ,$ here $\sigma _{y}$ is the second
Pauli matrix and the conjugation occurs in the computational basis $%
(|00\rangle $, $|01\rangle $, $|10\rangle $, $|11\rangle )$.
$C_{\rho }\left( t\right) $ quantifies the amount of quantum
correlation that is present in the system and can assume values
between $0$ (only classical correlations) and $1$ (maximal
entanglement).
\begin{figure}[tbph]
\begin{center}
\includegraphics[width=11cm,height=5cm]{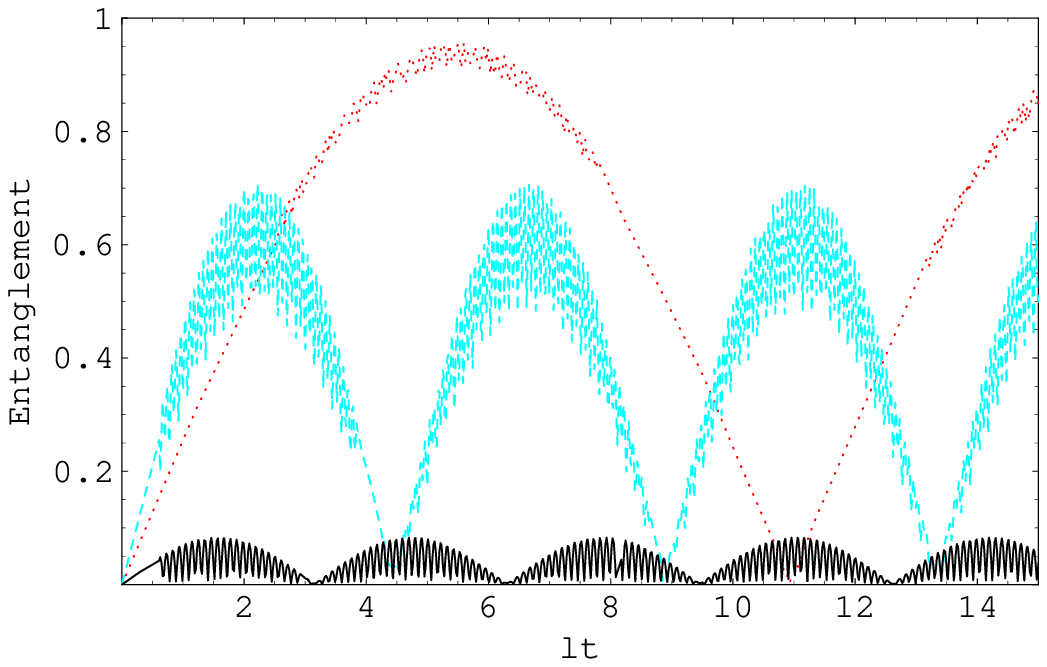} %
\includegraphics[width=11cm,height=5cm]{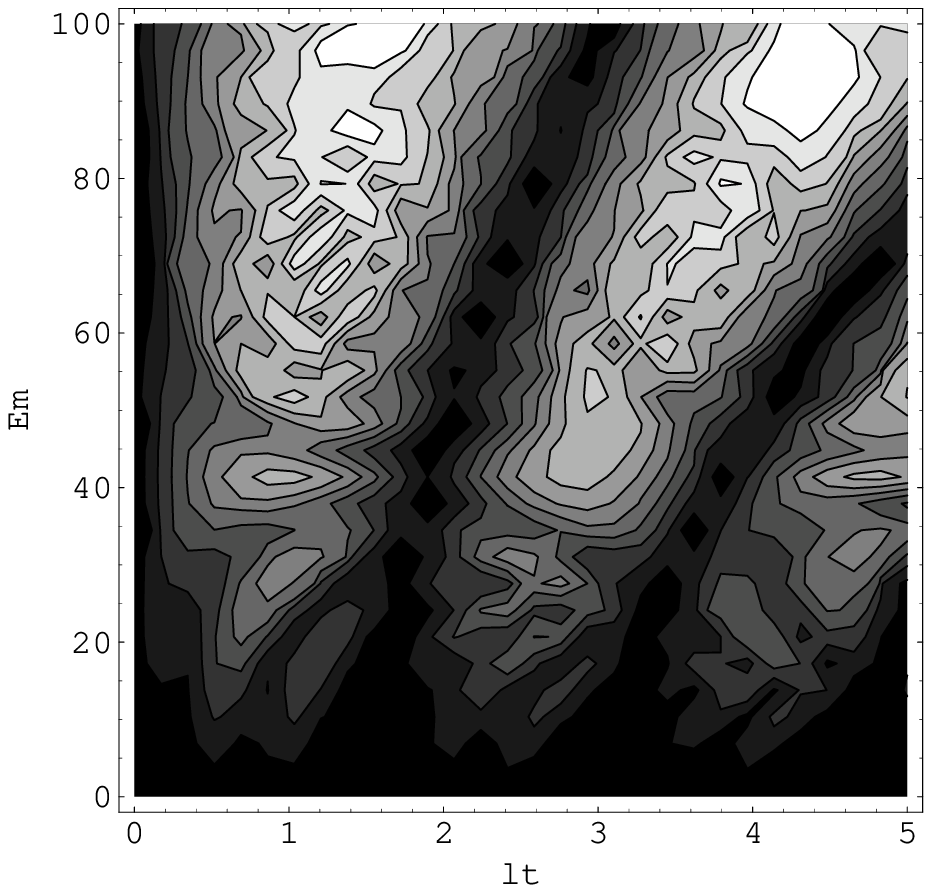}
\end{center}
\caption{{}The time evolution of the concurrence as a function of
the scaled
time $\protect\lambda t.$ The parameters are \emph{E}$_{\emph{J1}}$\emph{=30}%
$\protect\mu $eV , \ \emph{E}$_{\emph{J2}}$\emph{=2}$\protect\mu $eV, \ $%
\protect\gamma =0$ and different values of the coupling energy, where \emph{E%
}$_{m}$\emph{=200}$\protect\mu $eV (dotted curve), \emph{E}$_{m}$\emph{=60}$%
\protect\mu $eV (dashed curve) and
\emph{E}$_{m}$\emph{=5}$\protect\mu $eV (solid curve). Regions of
the entanglement sudden death are painted in gray.} \label{4}
\end{figure}

In figure \ref{4}, we plot the concurrence as a function of the
scaled time assuming that the two-coupled superconducting charge
qubits start from their excited states. The maximum value of the
entanglement decreases as the coupling energy is decreased. As
time goes on, the entanglement reaches zero value in a periodic
way, this period decreases as the coupling energy decreases. In
the uncoupled situation each qubit oscillates with its own
frequency and $C_{\rho }\left( t\right) =0$. It is interesting to
note that, the maximum entanglement is achieved at specific
choices of the interaction time i.e. the entanglement content
corresponding to specific choices of the interaction time and
large values of the coupling energy. We therefore consider the
question of how the coupling energy affect the entanglement of
system. In relation to that discussion, it is useful to examine
the effect of the characteristic energies with fixing the coupling
energy of the two charge qubits that change their state during the
transition. That criterion is related to, but clearly distinct
from, the question of quantifying how maximum a quantum state is.
Also, as can be seen from the graphs figures \ref{2} and \ref{4},
the entanglement vanishes at the time at which the population of
the symmetric state is maximal.
\begin{figure}[tbph]
\begin{center}
\includegraphics[width=11cm,height=5cm]{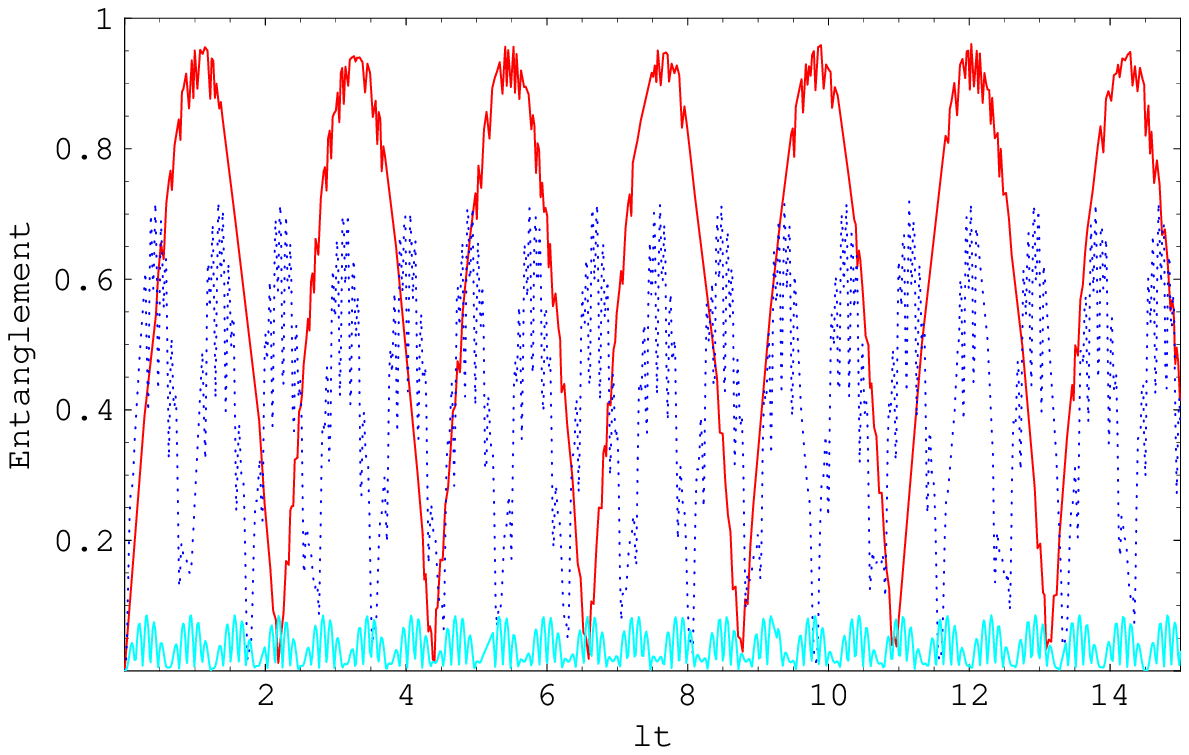} %
\includegraphics[width=11cm,height=5cm]{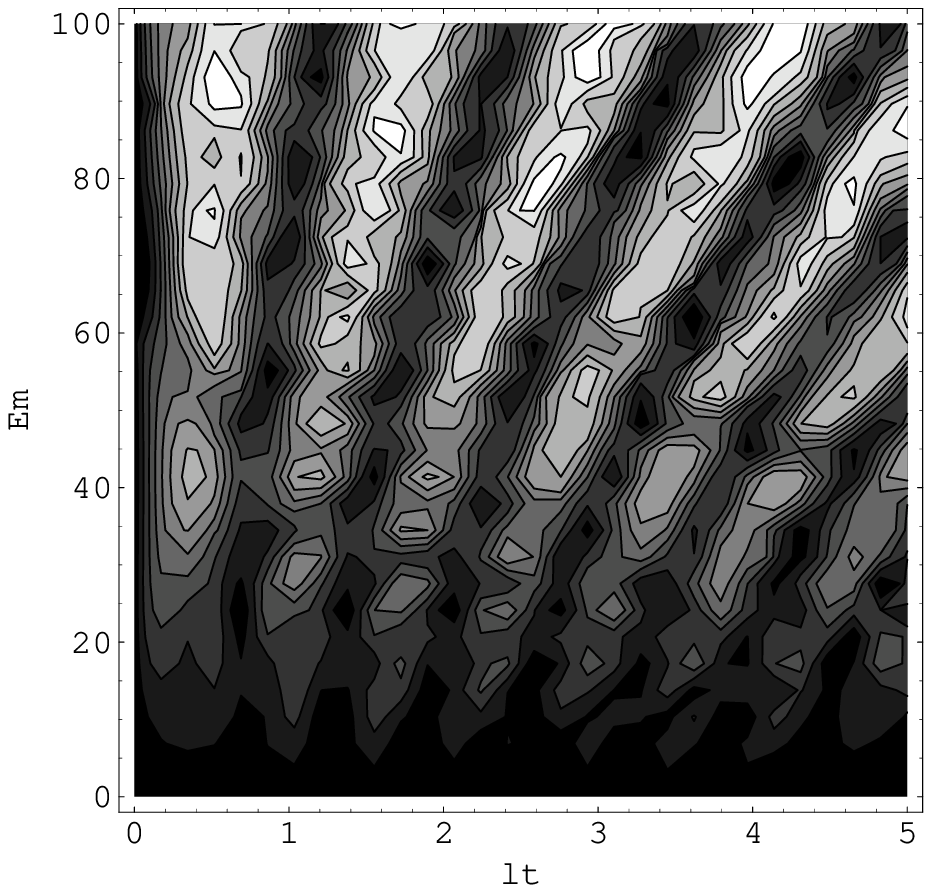}
\end{center}
\caption{{}{}The same as figure 4 but \emph{E}$_{\emph{J2}}$\emph{=5}$%
\protect\mu $eV.} \label{5}
\end{figure}

Put differently, with a small difference between the
characteristic energies
of the Cooper pairs (Josephson energy of the second qubit, \emph{E}$_{\emph{%
J2}}$\emph{=5}$\mu $eV) and still kept the parameter values of
figure \ref{4}, we show that the entanglement features are visibly
worsened, as in this regime oscillations are faster than the
previous case (see figure \ref{5}) in which big difference between
the characteristic energies of both qubits are considered. One has
still zero entanglement due to the time development, but with a
short period of the interaction time which is roughly given by
$\lambda t\approx \frac{2n\pi }{3},(n=1,2,3,...)$. Similar to the
previous case, one will have a very small amount of entanglement
when the coupling energy decreases and this amount disappear
completely when the coupling energy decreases further. Indeed, the
comparison of plots figure \ref{4} and figure \ref{5},
demonstrates that the entanglement in both cases has somewhat
similar behavior corresponding to different values of coupling
energy. The deviation value of the Josephson energies of the
Cooper pairs effect on the entanglement is particularly pronounced
as this deviation is much bigger.

Conventionally, maximally entangled state emerge from the coupling
of the two qubits by a small island overlapping both Cooper pair
boxes, i.e. two-coupled superconducting charge qubits. As such,
the phenomenon is the result of many-particle dynamics, often
described by a simple interaction model. In the single-qubit case
novel features appear which are due to the coherent microscopic
dynamics. Our study allows us to identify the dependence of these
features on different parameters of the system, thereby giving us
insight into how maximally entangled state and hence maximum
entanglement arise from dynamics in this particular coupling
process.

The quantum features of many systems decay uniformly as the result
of decoherence and much effort has been directed to extend the
coherence time of these qubits. However, it has been shown that
under particular circumstances where there is even only a partial
loss of coherence of each qubit, entanglement can be suddenly and
completely lost \cite{ebe07,alm07}.

\begin{figure}[tbph]
\begin{center}
\includegraphics[width=11cm,height=5cm]{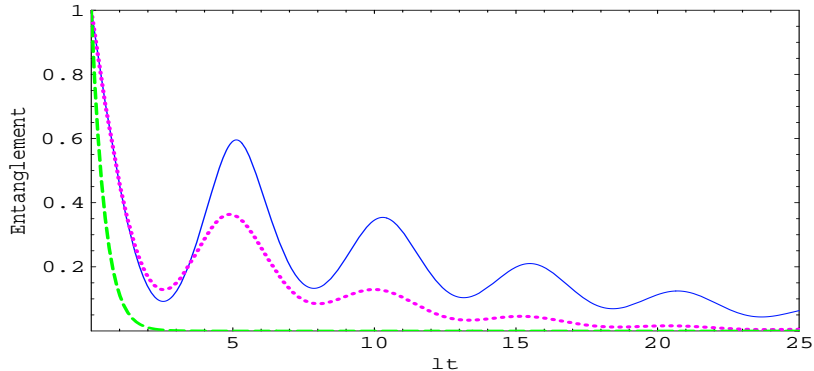}
\end{center}
\caption{{}{}{}The same as figure 5 but \emph{E}$_{m}$\emph{=200}$\protect%
\mu $ and different values of the decoherence parameter, where $\protect%
\gamma =0.01$ (solid curve), $\protect\gamma =0.1$ (dotted curve) and $%
\protect\gamma =0.8$ (dotted-dashed curve).} \label{6}
\end{figure}
This has motivated us to consider the question of how decoherence
effects the scale of entanglement in the present system. The
decoherence time due to the coupling to the vacuum via the probe
junction can be estimated to be roughly $100$ $ns$ at the resonant
condition. For the most of the experimental devices \cite{pas03},
this probing time restricted the upper limit of the decoherence
time, since the probe junction attached to the box had a sampling
time of typically $8$ $ns$. Once, the environment has been
switched on, i.e., $\gamma \neq 0$, it is very clear that the
decoherence plays a usual role in destroying the entanglement. In
this case and for different values of the decoherence parameter
$\gamma ,$ we can see from figure (\ref{6}) that after the onset
of the interaction the entanglement function increases to reach
its maximum showing strong entanglement. However its value
decreases after a short period of the interaction time to reach
its minimum. The function starts to increase its value again
however with lower local maximum values showing a strong decay as
time goes on. It is interesting to remark that decoherence due to
normal decay is often said to be the most efficient effect in
physics. Which means that, the entanglement increases rapidly,
then approaches to a minimum value in a periodic manner. Also,
from numerical results we note that with the increase of the
parameter $\gamma $, a rapid decrease of the entanglement
(entanglement sudden death) is shown \cite{yu06}.

Since the discovery of entangled sudden death \cite{yu06,yu04}, a
large number of instances of this surprising effect have been
identified in the theoretical literature
\cite{fic06,der06,pin07,ann07}. In general, decay takes infinitely
long, so one can wait any length of time. However, if the
entanglement reaches zero in a finite time \cite{yu06,yu04} the
game is over. Thereafter no distillation process exists that will
recover any useful feature of entangled quantum joint coherence
for use in quantum computing or communication \cite{ebe07b}. These
results should mark an important consideration in the design and
operation of future quantum information networks. Also, properties
of such entanglement decay depends on the coupling energy and
qubits Josephson energies. Even if achieving the maximally
entangled state is not possible in the presence of the
decoherence, one can argue that finding a maximally entangled
state can, under certain conditions, be achieved for a short
interaction time. It is also worth noting here that we have used
the simple model of a two-Cooper pair box problem, which
represents a proper physical system that can be used as a qubit.
This system is of a great interest because it offers the
possibility of scaling to a large number of interacting qubits.

\section{Conclusion}

In the important context of quantum state engineering and
characterization, we have studied the entanglement properties of
the special class of two-coupled superconducting charge qubits.
Clear physical interpretations for the maximally entangled state
generation and entanglement found for certain parameter regimes of
the system have been provided. In a strong coupling regime, two
different types of maximally entangled states of the system have
been created. This helps the comprehension of the quantitative
results achieved by the use of the concurrence. Moreover, we find
peculiar entanglement characteristics which are unique to this
system, and which we trace back to the interplay of the various
time scales of the dynamics. Our results suggest that the
maximally entangled mixed state is generic in different situations
for the coupled-charge qubits system, and that developing
entanglement theory under other sorts of restrictions is a
promising direction for further study. In a more general context,
our results provide further insight into the coupled dynamics of
superconducting charge qubits in the spirit of the experimental
realization \cite{gri07,pas03,liu06} and may provide a useful
maximally entangled states source in the exploration of various
quantum-information processing.

\[
\]


\end{document}